\documentclass[%
 reprint,
 amsmath,amssymb,
 aps,
]{revtex4-1}

\usepackage{graphicx}
\usepackage{subfigure}
\usepackage{dcolumn}
\usepackage{bm}
\usepackage[flushleft]{threeparttable}
\usepackage{url}
\usepackage{amsmath}
\usepackage{amssymb}
\usepackage{braket}
\usepackage{multirow}
\usepackage{array}
\usepackage{url}
\usepackage{booktabs}
\usepackage{xcolor}
\usepackage{framed}
\usepackage{lipsum}
\usepackage{color,soul}
\usepackage{lineno}
\colorlet{shadecolor}{blue!60}
\newcolumntype{P}[1]{>{\centering\arraybackslash}p{#1}}

\begin{document}

\preprint{APS/123-QED}
\title{Widespread negative longitudinal piezoelectric responses in ferroelectric crystals with layered structures}

\author{Yubo Qi$^1$ and Andrew M. Rappe$^2$}

\affiliation{%
$^1$ Department of Physics $\&$ Astronomy, Rutgers University, \\
Piscataway, New Jersey 08854, United States \\
$^2$ Department of Chemistry, University of Pennsylvania, \\ 
 Philadelphia, PA 19104-6323, United States\\
 }%

\begin{abstract}

In this study, we investigate the underlying mechanisms of the universal negative piezoelectricity in low-dimensional layered materials by carrying out first-principles calculations.
Two-dimensional layered ferroelectric CuInP$_2$S$_6$ is analyzed in detail as a typical example,
but the theory can be applied to any other low-dimensional layered piezoelectrics.
Consistent with the theory proposed in [\href{https://journals.aps.org/prl/abstract/10.1103/PhysRevLett.119.207601}{Physical Review Letters \textbf{119}, 207601 (2017)}],
the anomalous negative piezoelectricity in CuInP$_2$S$_6$ also results from its negative clamped-ion term,
which cannot be compensated by the positive internal strain part.
Here, we focus on a more general rule by proposing that having a negative clamped-ion term should be universal among piezoelectric materials, which is attributed to the ``lag of Wannier center" effect.
The internal-strain term, which is the change in polarization due to structural relaxation in response to strain, is mostly determined by the spatial structure and chemical bonding of the material.
In a low-dimensional layered piezoelectric material such as CuInP$_2$S$_6$, the internal-strain term is approximately zero.
This is because the internal structure of the molecular layers, which are bonded by the weak van der Waals interaction, 
responds little to the strain. 
As a result, the magnitude of the dipole, which depends strongly on the dimension and structure of the molecular layer, also has a small response with respect to strain. 
An equation bridging the internal strain responses in low-dimensional and three-dimensional piezoelectrics is also derived to analytically express this point.
This work aims to deepen our understanding about this anomalous piezoelectric effect, especially in low-dimensional layered materials, and provide strategies for discovering materials with novel electromechanical properties.
  
\end{abstract}

\pacs{Valid PACS appear here}

\maketitle

Piezoelectrics are a family of materials which enable interconversion between electrical energy and mechanical energy,
offering a wide range of applications, such as pressure sensors, actuators and noise attenuators~\cite{Martin72p1607,Yang09p34}.
The piezoelectric tensor is expressed as the change of the polarization with respect to a strain.
The piezoelectric coefficients are usually positive, indicating that the polarization is more likely to increase under a tensile strain~\cite{Katsouras16p78}.
However, researchers in recent years have seen some exceptions, such as a variety of $ABC$ ferroelectrics~\cite{Liu17p207601}
and several \uppercase\expandafter{\romannumeral3}-\uppercase\expandafter{\romannumeral5} zincblende compounds~\cite{Bellaiche00p7877}.
The piezoelectric coefficient can be decomposed into a clamped-ion term and an internal-strain term~\cite{Saghi98p4321,Saghi99p12771,Corso1994p10715,Bellaiche00p7877,Liu17p207601,Liu20p197601}.
A recent theoretical work by Liu and Cohen~\cite{Liu17p207601} clearly revealed the origin of negative piezoelectricity, which results from the domination of the negative clamped-ion term over the internal strain term.
This theory inspires us to investigate a more general question `whether negative clamped-ion terms are universal among all piezoelectrics'. And if so, what is the underlying physics of this rule?
Moreover, Liu and Cohen's work also demonstrated that the signs of piezoelectric responses are determined by the competition between the clamped-ion and internal strain terms.
Here, we are particularly interested in identifying a family of piezoelectrics tending to have a smaller internal strain term, which cannot overcome the clamped-ion part.

In this work, we addressed all these questions by investigating the consistently negative piezoelectric responses in low-dimensional layered materials.
We note that nearly all the low-dimensional layered piezeoelectrics reported so far~\cite{Liu20p197601}, such as polyvinylidene fluoride (PVDF) and its copolymers~\cite{Katsouras16p78}, 
CuInP$_2$S$_6$ (CIPS)~\cite{Neumayer19p024401,You19p3780,Brehm20p43}, and bismuth tellurohalides~\cite{Kim19p104115}, exhibit negative longitudinal piezoelectricity.
Here, we select CIPS as an example and perform density functional theory (DFT) calculations [See Supplementary Materials (SM) section \uppercase\expandafter{\romannumeral1}  computational details], 
but the rules acquired can be applied to any other low-dimensional layered materials.
Similar to the three-dimensional piezoelectrics with negative piezoelectric responses~\cite{Liu17p207601}, CIPS also has a small internal-strain term, which can not compensate the negative clamped-ion term. 
Moreover, Ref.~\cite{Kim19p104115} demonstrated that bismuth tellurohalides, another kind of two dimensional piezoelectrics,
also exhibit negative piezoelectric responses majorly resulting from charge redistribution under stress,   
providing another strong example of this physical scenario.
In this work, we focus on the general aspects of the two competing piezoelectric terms.
We demonstrate that most piezoelectrics should have a negative clamped-ion term.
This is because as the volume expands with atomic fractional coordinates remaining fixed,
chemical bond lengths elongate homogeneously, but the Wannier centers cannot follow this homogeneous strain.  
As a result, polarization decreases. 
We refer to this phenomenon as the ``lag of Wannier center'' effect~\cite{Bellaiche00p7877}.
Moreover, not only in CIPS, the internal-strain response is expected to be tiny among almost all the low-dimensional layered piezoelectrics.
This effect is attributed to the fact that the inter-layer van der Waals (vdW) bonding is much weaker than the intra-layer chemical bonding. 
Under stress, the inter-layer gap will take most of the change, making the dimension of a molecular layer and the dipole associated with it change very little.
To better illustrate this point, we derive an analytical expression demonstrating the difference and correlation between the internal strain responses in three-dimensional and low-dimensional piezoelectrics.
Our result shows that the responses of internal coordinates with strain in a low-dimensional layered material is about one decade smaller than that in a conventional three-dimensional piezoelectric.
These analyses successfully explain why negative piezoelectricity is expected to be widespread in low-dimensional layered materials.

\begin{figure}[htbp]
\includegraphics[width=7.5cm]{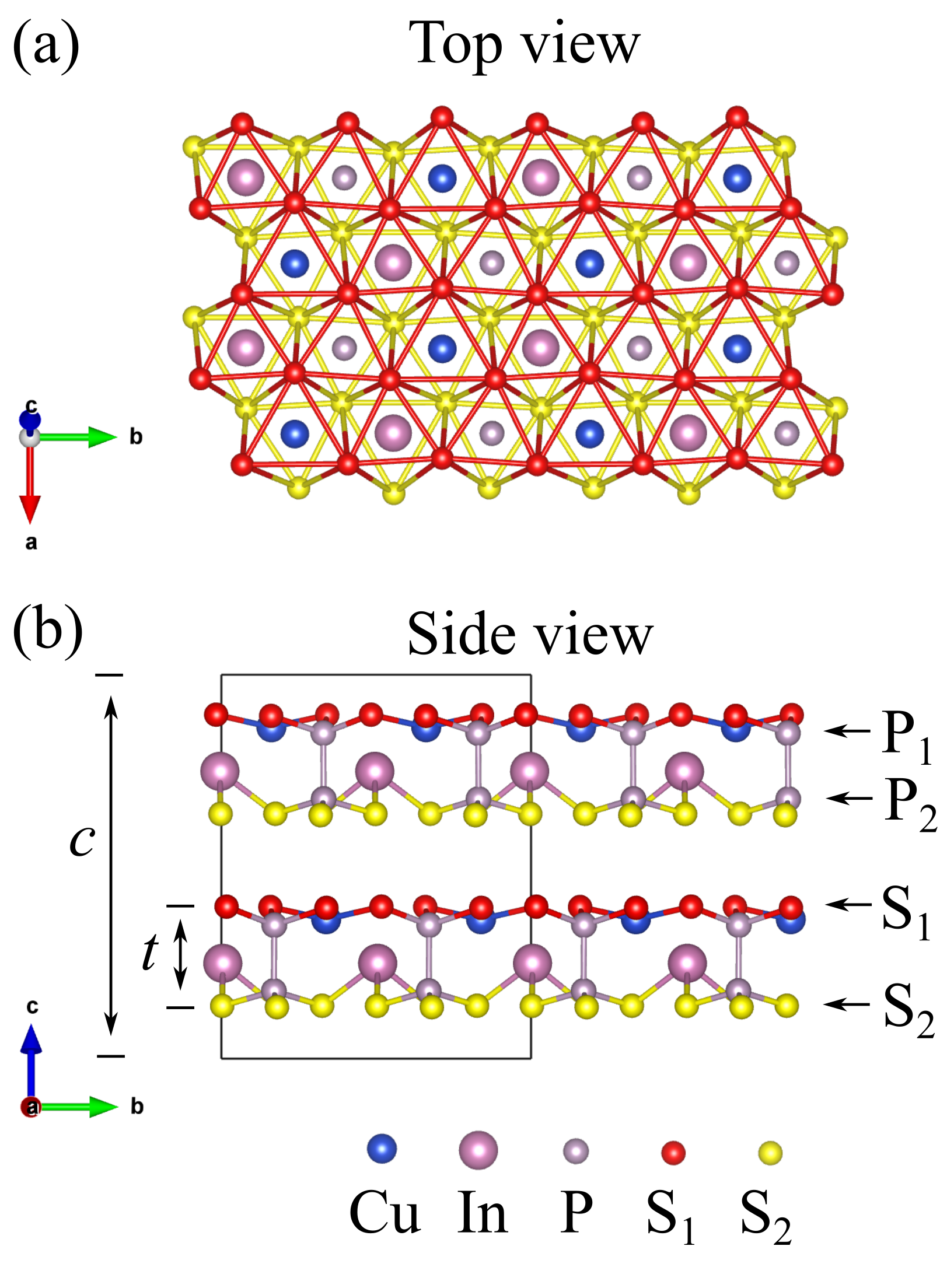}
\caption{Structure of CuInP$_2$S$_6$ (CIPS). The S atoms in the top and bottom layers are represented with different colors (red and yellow) and labels (S$_1$ and S$_2$).
(a) Top view of ferroelectric CIPS. Each sulfur octahedron is filled with a Cu, In atom or a P-P dimer; 
(b) side view of ferroelectric CIPS, with lattice parameters $a=6.10$ \AA, $b=10.56$ \AA, and $c=13.62$ \AA. 
The upper and lower P atoms are labeled as P$_1$ and P$_2$. 
$t$ molecular-layer thickness and $c$ is the height of the cell.}\label{f1}
\end{figure}

CIPS is a two-dimensional (2D) ferroelectric material composed of polar molecular layers held together with weak vdW forces~\cite{Liu16p12357,Maisonneuve95p157,Maisonneuve97p10860}.
Sulfur octahedra form the framework of a molecular layer, and each sulfur octahedron is filled with a Cu atom, In atom, or a P-P dimer [Fig.~\ref{f1} (a)].
Since Cu and P-P dimer exchange their sites in adjacent molecular layers, each primitive cell is composed of two molecular layers. 
In each Cu-filled sulfur octahedron, the Cu atom has two different possible occupation sites, above and below the center plane, corresponding to two polar states.
At low temperature, CIPS adopts its ferroelectric phase, with all or most Cu atoms displaced in the same direction, as shown in Fig.~\ref{f1} (b).
Above its Curie temperature $T_C$ ($\approx315$ K)~\cite{Maisonneuve97p10860}, 
CIPS becomes paraelectric due to the equal up or down site occupancy by Cu atoms. 
Here, we should note that even though the In atoms are also displaced off-center, their displacements are far smaller than those of Cu [$d(\rm{In})=0.22$~\AA\ {\em{vs.}} $d(\rm{Cu})=1.28$~\AA]. 
Therefore, the ferroelectricity in CIPS mostly originates from the Cu displacements.
In this study, we focus on the intrinsic electromechanical properties of the ground-state ferroelectric CIPS.
Therefore, temperature-induced cation disorder is beyond the consideration of this work~\cite{Maisonneuve97p10860,Neumayer19p024401}.

\begin{figure}[htbp]
\includegraphics[width=7.5cm]{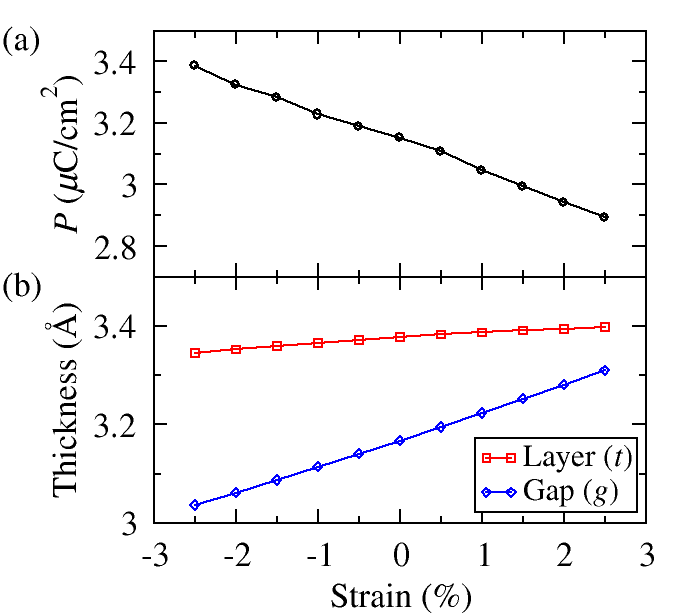}
\caption{
Changes of (a) polarization, (b) inter-layer gap length and layer thickness with respect to strain.}\label{f2}
\end{figure}

The optimized lattice constants obtained from our DFT calculations are $a=6.09$ \AA, $b=10.56$ \AA, and $c=13.76$ \AA, 
which match experimental ones ($a=6.10$ \AA,  $b=10.56$ \AA, $c=13.62$ \AA) very well~\cite{Maisonneuve97p10860}, with only 1.0\% error. 
Polarization, calculated via the Berry's phase method, is $P=3.04$ $\mu$C/cm$^2$, 
which is also consistent with the experimental values ($2.55\sim3.80$ $\mu$C/cm$^2$~\cite{Maisonneuve97p10860,Liu16p12357}). 
All of these results demonstrate the reliability of our first-principles calculations.
 In this study, we focus on the longitudinal component $e_{33}$ only. 
 Therefore, for simplicity, all the symbols of vectors or tensors (such as piezoelectric tensor $e$, polarization $P$, strain $S$, and lattice axis $c$) refer to the $z$ or $zz$ components, 
 unless specifically stated.
To evaluate the piezoelectric coefficient $e=\left({\partial{P}}/{\partial{S}}\right)_E$, where $P$ is the polarization, $S$ is the strain and $E$ is the electric field,
we artificially change the lattice parameter $c$, which is also the height of the primitive cell, with the in-plane lattice parameters fixed, relax the structure, and then calculate the polarization.
As shown in Fig.~\ref{f2} (a), the polarization decreases with increasing tensile strain, indicating a negative piezoelectric coefficient.
The longitudinal piezoelectric coefficient $e$ is $-9.6$ $\mu$C/cm$^2$ from our DFT calculations.

The piezoelectric coefficient can be decomposed into two parts as~\cite{Saghi98p4321,Saghi99p12771,Corso1994p10715,Bellaiche00p7877},
\begin{equation}
e=\frac{\partial{P}}{\partial{S}}=e^{\left(0\right)}+e_{i},
\end{equation}
where $e^{\left(0\right)}$ is the clamped-ion term, and
\begin{equation}
e_{i}=\sum_n\frac{qc}{\Omega}Z^*\left(n\right)\frac{\partial{U}\left(n\right)}{\partial{S}}
\end{equation}
is the internal-strain part [See SM Section \uppercase\expandafter{\romannumeral2} for schematic illustrations of the two terms].
$U$ is the fractional atomic coordinates in the supercell and $n$ runs over all atoms in a unit cell. 
Here,  
$\Omega$ is the volume of a unit cell, $q$ is the electronic charge, and $Z^*$ is the Born effective charge.
The clamped-ion response is evaluated at zero internal strain, which means that the internal fractional coordinates are frozen,
and reflects the redistribution of electrons with respect to a homogeneous strain.
It features the change of Born effective charges, since the polarization, which is expressed as dipole over volume, remains the same under a homogeneous strain and fixed Born effective charges.
On the other hand, the internal strain term describes the internal distortion under a macroscopic strain, assuming the Born effective charges fixed.
The values of the clamped-ion and internal strain terms are summarized in Table~\ref{t1}, 
from which we can see that the negative piezoelectricity in CIPS almost completely originates from the negative clamped-ion term.
Actually, having a negative clamped-ion piezoelectric response is universal among piezoelectrics
(See SM Section \uppercase\expandafter{\romannumeral3} for the previously reported clamped-ion terms in other piezoelectrics), 
which means that the Wannier centers generally 
fail to follow anions (in fractional coordinates) fully upon a tensile strain~\cite{Bellaiche00p7877}. 
This ``lag of Wannier center'' results from the damping of Coulombic repulsion between electrons as the cation and anion separate, 
which will be discussed in more details in the following subsection. 
It is worth mentioning that, according to our DFT calculations, the in-plane lattice constants change very little with the $c$ lattice, suggesting a nearly zero Poisson’s ratio of CIPS (less than 0.05).
Therefore, a longitudinal strain causes little in-plane deformation or improper piezoelectric effect.

To explain this lag of Wannier center effect, we begin by discussing the ionicity of a chemical bond.
Ionicity describes the extent of electron gain in the anion and may have different mathematical expressions.
In the Coulson model~\cite{Abu07p085201,Coulson62p357},
the bond connecting atoms $A$ and $B$ is expressed with the linear combination of atomic orbitals (LCAO) approximation as
\begin{equation}
\psi=c_A\phi_A+c_B\phi_B,
\end{equation}
where $\phi_A$ and $\phi_B$ are atomic orbitals centered on atoms $A$ and $B$.
The ionicity $I_C$  is expressed as
\begin{equation}
I_C=\frac{c_A^2-c_B^2}{c_A^2+c_B^2}.
\end{equation}
In the model based on the centers of maximally localized Wannier functions~\cite{Abu07p085201,Marzari97p12847},
The ionicity $I_W$ is given as
\begin{equation}
I_W=\left(2\beta-1\right)^{N/M},
\end{equation}
where $N$ is the atomic valency, $M$ is the coordination number, and $\beta=r_w/d_1$.
$r_w$ is the distance between the Wannier center and the position of cation and
$d_1$ is the bond length (Fig. S2). Therefore,
$\beta-0.5$ describes the deviation of the Wannier center from the bond center.
Here, we should emphasize that even though the expressions are different in the two models, they gauge the same physical quantity and give comparable values~\cite{Pilania15p26}.

Here, we consider a linear $\cdots{A}$--$B$--${A}$--${B}\cdots$ atomic chain, with alternative unequal bond lengths $d_1$ and $d_2$ ($d_2>d_1$).
The length of a unit cell is $d=d_1+d_2$.
In a two-atom-basis scheme~\cite{Nielsen12p114304}, the Bloch state of the an electron with the wavenumber ${{k}}$ can be expressed as
\begin{equation}
\begin{split}
&\psi_{{k}}\left({{r}}\right)=\frac{1}{\sqrt{N}}\sum_{{R}}e^{i{kR}} \\
&\left[c{_{A,{k}}}\phi_{A}\left({r-R_A-R}\right)+c{_{B,{k}}}\phi{_B}\left({r-R_B-R}\right)\right].\\
\end{split}
\end{equation}
However, this expression is based on an one-electron model and neglects the electron-electron interaction, which plays an important role in the ``lag of Wannier center" effect.
Here, we assume that each atomic pair $A$--$B$ contributes two electrons (denoted as $i$ and $j$) to the valence band.
Considering the electron-electron interaction~\cite{Fuchs08p11}, the Bloch state should be modified to
\begin{equation}
\Psi_{{k}}\left({{r}}\right)=\psi_{{k}}\left({{i,r}}\right)\psi_{{k}}\left({{j,r}}\right).
\end{equation}
Here, $\psi{_{{k}}}\left(i\right)$ is the wavefunction of the electron $i$ whose expression is the same as equation (6).
The Hamiltonian is expressed as~\cite{Klopman64p4550} 
\begin{equation}
\hat{H}=\hat{K}_e-\frac{1}{N}\sum_{R}\left[\frac{Z_A}{r_{iAR}}+\frac{Z_B}{r_{iBR}}+\frac{Z_A}{r_{jAR}}+\frac{Z_B}{r_{jBR}}\right]+\frac{1}{r_{ij}},
\end{equation}
where $\hat{K}_e$ is the kinetic energy, $A$ and $B$ represent the atoms, 
$Z_A$ and $Z_B$ are the effective nuclear charges,
and $i$ and $j$ correspond to the electrons.
$\frac{Z_A}{r_{iAR}}\equiv\frac{Z_A}{{r_i\left(r-R_A-R\right)}}$ is the interaction between the electron $i$ and the nuclear of atom $A$ (whose position is $R_A$ in a unit cell) in the unit cell locating at $R$. 
We neglect the nuclear-nuclear interaction, which should be a constant regardless of electron distribution.
The energy can be approximated as
\begin{equation}
\begin{split}
 E_k&=2\left(c_{A,k}^2K_A+c_{B,k}^2K_B\right)-2c_{A,k}^2\left(E_A+\Gamma_{Ba}\right)\\
& -2c_{B,k}^2\left(E_B+\Gamma_{Ab}\right)-4c_{A,k}c_{B,k}\left(\Gamma_{AB}+\cos{k}\Gamma_{AB}^*\right)\\
 &+c_{A,k}^4\Gamma_{aa}+c_{B,k}^4\Gamma_{bb}+2c_{A,k}^2c_{B,k}^2\left(\Gamma_{ab}+\cos{k}\Gamma_{ab}^*\right)
\end{split}
\end{equation}
Here, $K_{A(B)}$ is the kinetic energy of the electron on atomic orbital $\phi_{A(B)}$,
$E_{A(B)}$ is an atomic term describing the core-electron interaction inside an atom, 
$\Gamma_{Ba}$ and $\Gamma_{Ab}$ describe the core-electron interaction between atoms, 
$\Gamma_{AB}$ is the intra-cell resonance term,
$\Gamma_{AB}^*$ in the inter-cell resonance term,
$\Gamma_{aa(bb)}$ is another atomic term describing the electronic Coulomb repulsion,
$\Gamma_{ab}$ 
describes the Coulomb repulsion between two electrons belong to the two atomic orbitals in a unit cell,
and 
$\Gamma_{ab}^*$
describes the Coulomb repulsion between two electrons belong to the two atomic orbitals in neighboring unit cells
[See SM Section \uppercase\expandafter{\romannumeral4} for the derivation of the equation (8)  and the expression of each term].
The Coulombic terms $\left(\Gamma_{ab}+\cos{k}\Gamma_{ab}^*\right)$ favor a large difference between $c_A$ and $c_B$, which means a large $I_C$ and Wannier center far away from the bond center.
As the bond elongates, the electronic-wavefunctions overlap $\left|\phi_A(i,r)\right|^2\left|\phi_B(j,r-d_1)\right|^2$ and $\left|\phi_A(i,r)\right|^2\left|\phi_B(j,r-d_1+d)\right|^2$ in the Coulombic terms (See SM equations S15 and S16) reduce, and the magnitude of $\left(\Gamma_{ab}+\cos{k}\Gamma_{ab}^*\right)$ and $\left|c_A-c_B\right|$ also decrease.
This means that the Wannier center stays closer to the center of the bond, rather than following the anion completely [Fig. S2 (b)].
Even though the resonance term works against the Coulombic term [Fig. S2 (c)], decrease of the electronic-wavefunctions overlap has a more profound influence on the latter [Fig. S2 (d)],
since it involves a near-site interaction $\frac{1}{r_{ij}}$ (See SM equations S15 and S16) in all space.
This ``lag of Wannier center'' effect provides two deductions. First, the ionicity tends to decrease with increasing bond length, which is consistent with previous study~\cite{Pilania15p26}.
Besides, the absolute values of the Born effective charges should also decrease with increasing bond lengths (or upon a tensile strain), which conforms to this CIPS case, as shown in Table~\ref{t1}.

The internal-strain contribution to piezoelectricity in CIPS is positive, but not big enough to neutralize the negative clamped-ion term.
The small internal-strain term is mainly attributed to the low dimensionality of CIPS.
In low-dimensional layered materials, the inter-layer vdW interaction is much weaker than the intra-layer chemical bonding.
As a result, the inter-layer gap will take most of the change in the dimension of the cell when it is stressed.
In Fig.~\ref{f2} (b), we plot the changes of molecular-layer thickness $t$ and inter-layer gap $g$ lengths under various strains from DFT calculations.
As expected, $g$ grows much faster than $t$ under tensile strain.
This also means that the ratio $R$ of a molecular layer thickness $t$ to the lattice $c$ of the cell decreases with the strain, which is expressed as 
\begin{equation}
R{_S}=\frac{\partial{R}}{\partial{S}}<0.
\end{equation}
This negative $R_S$ leads to a small $U_S\ {\equiv}\ {\partial{U}}/{\partial{S}}$.
To illustrate this point, we begin with treating each molecular layer as a free-standing crystal
and investigate how the atomic fractional coordinate within the layer $u$ changes with the strain of the molecular layer $s$.
In Table~\ref{t1}, we list the values of $u_s\ {\equiv}\ {\partial{u}}/{\partial{s}}$ for all four atom types in CIPS. 
The values of $u_s$ in two typical three-dimensional piezoelectrics (in which $U_S=u_s$) BaTiO$_3$ and PbTiO$_3$ are also listed for comparison.
We can see that the $u_s$ in CIPS are in the same order of magnitude with the $U_s=u_s$ in BaTiO$_3$ and PbTiO$_3$, indicating that 
there is little difference between the piezoelectric property of a single molecular layer and those of conventional three-dimensional piezoelectrics.
To understand the origin of the small cell-scale response $U_S$ in CIPS, 
we derive the conversion formula between $u_s$ and $U_S$  [See SM Section \uppercase\expandafter{\romannumeral6} for details of this derivation], 
which is expressed as
\begin{equation}
U_S=\left(R+R_S\right)u_s+R_Su.
\end{equation}
The coefficient in the first term $\left(R+R_S\right)$ is the scaling factor. 
Since $R$ is less than 1 and $R_S$ is negative, $U_S$ is expected to be much smaller than $u_s$.
In addition to the layer-scale fractional coordinate $u$ changes,
the change of the ratio between layer thickness and cell lattice also affects the magnitude of polarization.
This effect is described by the second term, and its contribution is also negative. 
In Table~\ref{t1}, we list the contributions from the two terms and find that $U_S$ is approximately one order of magnitude less than $u_s$. 

\begin{table*}
\begin{tabular}{ P{3.0cm}  P{2.0cm} P{2.0cm} P{2.0cm}  P{2.0cm} P{2.0cm}P{2.0cm}}
\hline
\hline
\multicolumn{7}{c}{CuInP$_2$S$_6$  }   \\
\hline
\multicolumn{7}{c}{$e=-9.7$ \ \ \ \ \ \ \ \ $e_{i}$=0.4 \ \ \ \ \ \ \ \  $e^{\left(0\right)}=-10.1$ \ \ \ \ \ \ \ \ $R=0.2579$ \ \ \ \ \ \ \ \ $R_{S}=-0.1766$}   \\
\hline
Atom & Cu & In & P$_1$ & P$_2$ & S$_1$ & S$_2$ \\
\hline
$u$ &    0.3777  &  -0.0671 & -0.3463  & 0.3186  & 0.5000  & -0.5000 \\
$u_{s}$ & 0.5680  &  0.0697 &  0.1546  & -0.1732  & 0.0000  & 0.0000 \\
$u_{s}R$       & 0.1465  &  0.0180 &  0.0399  & -0.0447  & 0.0000  & 0.0000 \\
$u_{s}R_S$ & -0.1003 &  -0.0123 &  -0.0273  & 0.0306  & 0.0000  & 0.0000 \\
$u_{s}(R+R_S)$ &  0.0462 &   0.0057 &   0.0126  & 0.0141  & 0.0000  & 0.0000 \\
$uR_S$ & -0.0667  &  0.0119 &  0.0612 & -0.0563  & -0.0883  & 0.0883 \\           
$U_{S}$ & -0.0205 & 0.0175 & 0.0737 & -0.0704 & -0.0883 & 0.0883 \\
$Z^*$  & 0.6204 & 2.2432 & 0.9380 & 0.8416 & -0.8350 & -0.7156 \\
$\partial{\left|Z^*\right|}/\partial{S}$  & -2.6634 & -2.9360 & -0.5171 & -2.5774 & -1.5968 & -1.1126 \\
\hline
\hline
\end{tabular}
 
\begin{tabular}{ P{0.5cm} P{3.2cm} P{3.2cm} P{1.0cm}  P{3.2cm} P{3.2cm}  P{0.5cm}}
\\
\hline
\hline
&  \multicolumn{2}{c}{BaTiO$_3$}  & & \multicolumn{2}{c}{PbTiO$_3$}  & \\
\cline{2-3}  \cline{5-6}
& ${U_S(\rm{Ba})}={u_s(\rm{Ba})}$    &  ${U_S(\rm{Ti})}={u_s(\rm{Ti})}$     &  & ${U_S(\rm{Pb})}={u_s(\rm{Pb})}$      &  ${U_S(\rm{Ti})}={u_s(\rm{Ti})}$  &   \\
&  0.184  &  0.198  & & 0.279  & 0.151 & \\
\hline
\hline

 \end{tabular}
 \caption{Piezoelectricity and the contributions from different parts (the unit is $\mu$C/cm$^2$). 
The changes of cation displacements with strain in BaTiO$_3$ and PbTiO$_3$ are also listed for comparison.
}\label{t1}
\end{table*}

Another reason for the small internal strain term in CIPS, which plays a secondary role, is its small Born effective charges (Table~\ref{t1}).
In typical ABO$_3$ ferroelectric perovskites, whose Born effective charges are large,
the ferroelectricity results from the $p$-$d$ orbital hybridization induced Jahn-Teller distortion~\cite{Cohen92p136}.
The charge density distribution in these hybridized covalent bonds is sensitive to the change of cation displacement, indicating a large Born effective charge.
On the other hand, Cu and In atoms in CIPS make ionic bonds with the S octahedra.
Similar to the materials with geometric ionic size effect induced ferroelectricity, the Born effective charges in CIPS are small and close to the nominal ionic charges~\cite{Van04p164,Ederer04p849,Tohei09p144125}.

In summary, we investigate the negative piezoelectricity in low-dimensional layered materials by 
performing first-principles calculations.  
CuInP$_2$S$_6$ is selected as a typical example, but the theory is general and can be applied to any other low-dimensional layered piezoelectrics.
Consistent with the theory about the origin of negative piezoelectricity proposed by Liu and Cohen,
the negative piezoelectricity in CuInP$_2$S$_6$ also originates from a negative clamped-ion and an approximately zero internal strain term.
Furthermore, we emphasize that a negative clamped-ion piezoelectric response is universal among piezoelectrics, due to the ``lag of Wannier center'' effect.
In addition, the internal strain term is dramatically suppressed in low-dimensional layered piezoelectrics, 
since the thickness of a molecular layer and the dipole associated with it respond little to the strain state of the cell.
Based on these facts, we propose that negative piezoelectricity should exist widely in low-dimensional layered materials.
We hope that this work can provide more insight about the underlying physical mechanism in negative piezoelectricity and 
inspire the design of practical devices benefitting from materials with this novel electromechanical property.

\section*{ACKNOWLEDGMENTS}
We thank Liang Z. Tan for valuable discussions. 
This research is intellectually inspired by and primarily supported as part of the center for 3D Ferroelectric Microelectronics (3DFeM), an Energy Frontier Research Center funded by the U.S. Department of Energy (DOE), Office of Science, Basic Energy Sciences under Award Number DE-SC0021118,
including the final development of the analytic model of piezoelectricity. This research was initiated with the support of the U.S. Office of Naval Research, under grant N00014-20-1-2701, including the numerical computations and early variants of the model. A.M.R. designed the project, in consultation with Y.Q. Y.Q. conducted all calculations and physical analysis. Y.Q. drafted the manuscript, and all authors participated in rewriting.

\bibliography{rappecites}

\appendix

\end{document}